\documentclass[runningheads,a4paper]{llncs}

\usepackage{amsmath, amssymb} 
\usepackage{verbatim} 
\usepackage[pdftex]{graphicx}
\usepackage{setspace}
\usepackage{algorithmic}
\usepackage{algorithm} 
\usepackage{subfig}
\usepackage{times}
\usepackage{paralist}
\usepackage{cases}

\usepackage{paralist}
\usepackage{cases}

\usepackage{tabularx}

\usepackage{tikz}

\usepackage{color}

\usepackage{tikz}
\usetikzlibrary{patterns}
\usetikzlibrary{arrows}
\usepackage{sansmath}
\tikzset{
     task/.style={fill=#1,  rectangle},
     task1a/.style={task=green!30},
     task1b/.style={task=green},
     task2a/.style={task=orange!30},
     task2b/.style={task=orange},
     task3a/.style={task=pink},
 task3b/.style={task=pink!80},
     task4a/.style={task=cyan},
task4b/.style={task=cyan!50},
     task5/.style={task=blue},
     task6/.style={task=purple},
     task7/.style={draw,minimum height=\uy,},
     task8/.style={draw,thick},
     task9/.style={task=lightgray,draw,minimum height=\uy,},
     task10/.style={pattern=north west lines,draw,minimum height=\uy,},
}
\def\ux{0.6cm}\def\uy{0.5cm}

\newcommand{\setof}[1]{\left\{{#1}\right\}}

\newcommand{\kuframework}[1]{$\mathbf{k^2U}$}

\title{On Schedulability Analysis of EDF Scheduling by Considering
  Suspension as Blocking\thanks{This work has been supported by
    Deutsche Forschungsgemeinschaft (DFG), as part of Sus-Aware
    (Project no. 398602212) and the collaborative research center
    SFB876, subproject A1.}}

\author{Mario G\"unzel and Jian-Jia Chen} \institute{ Department of
  Informatics,
  TU Dortmund University, Germany\\
}
\begin{document}

\maketitle

\begin{abstract}
  During the execution of a job, it may suspend itself, i.e., its
  computation ceases to process until certain activities are complete
  to be resumed. This paper provides a counterexample of the
  schedulability analysis by Devi in Euromicro Conference on Real-Time
  Systems (ECRTS) in 2003, which is the only existing suspension-aware
  analysis specialized for uniprocessor systems when preemptive earliest-deadline-first (EDF) is applied
  for scheduling dynamic self-suspending tasks.
\end{abstract}

\section{Introduction}

Self-suspension behavior has been demonstrated to appear in complex
cyber-physical real-time systems, e.g., multiprocessor locking
protocols, computation offloading, and multicore resource sharing, as
demonstrated in \cite[Section 2]{Chen2018-suspension-review}.
Although the impact of self-suspension behavior has been investigated
since 1990, the literature of this research topic has been flawed as
reported in the review by
Chen~et~al.~\cite{Chen2018-suspension-review}.

Although the review by Chen~et~al.~\cite{Chen2018-suspension-review}
provides a comprehensive survey of the literature, two unresolved
issues are listed in the concluding remark. One of them is regarding
the ``\emph{correctness of Theorem 8 in \cite[Section
  4.5]{DBLP:conf/ecrts/Devi03} $\cdots$ supported with a rigorous
  proof, since self-suspension behavior has induced several
  non-trivial phenomena}''. This paper provides a counterexample of
Theorem 8 in \cite[Section 4.5]{DBLP:conf/ecrts/Devi03} and disproves
the schedulability test.

We consider a set of implicit-deadline periodic tasks, in which
each task $\tau_i$ has its period $T_i$, worst-case self-suspension
time $S_i$, and worst-case execution time $C_i$. The relative deadline
$D_i$ is set to $T_i$. There are two main models of self-suspending
tasks: the \emph{dynamic} self-suspension and \emph{segmented} (or
\emph{multi-segment}) self-suspension models. Devi's analysis in
\cite{DBLP:conf/ecrts/Devi03} considers the dynamic self-suspension
model. That is, a task instance (job) released by a task $\tau_i$ can
suspend arbitrarily as long as the total amount of suspension time of
the job is not more than $S_i$. 

The analysis by Devi in Theorem 8 in \cite[Section
4.5]{DBLP:conf/ecrts/Devi03} extended the analysis proposed by Jane
W.S. Liu in her book \cite[Page 164-165]{Liu:2000:RS:518501} for
uniprocessor preemptive fixed-priority scheduling to uniprocessor
preemptive EDF scheduling. Under preemptive EDF scheduling, the job
that has the earliest absolute deadline has the highest priority.
Despite the non-optimality of EDF for scheduling self-suspending task
systems as shown in
\cite{DBLP:conf/rtss/RidouardRC04,RTSS2016-suspension},
EDF remains one of the most adopted scheduling strategies.

Devi's analysis quantifies the additional interference due to
self-suspensions from the higher-priority jobs by setting up the
\emph{blocking time} induced by self-suspensions.  The correctness of
the analysis by Liu in~\cite[Page 164-165]{Liu:2000:RS:518501} has
been proved by Chen~et~al.~\cite{ChenECRTS2016-suspension} in 2016 for
fixed-priority scheduling.  The authors in
\cite{ChenECRTS2016-suspension} noted that ``\emph{Even though the
  authors in this paper are able to provide a proof to support the
  correctness, the authors are not able to provide any rationale
  behind this method which treats suspension time as blocking time.}''

Devi's analysis for implicit-deadline task systems is
rephrased as follows:
\begin{theorem}[Devi~\cite{DBLP:conf/ecrts/Devi03}]
\label{thm:devistheorem}
  Let $\textbf{T} = \setof{\tau_1, \tau_2, \ldots, \tau_n}$ be a
  system of $n$ implicit-deadline periodic tasks, arranged in order of
  non-decreasing periods. The task set $\textbf{T}$ is schedulable
  using preemptive EDF if 
\[
\forall k: 1 \leq k \leq n:: \frac{B_k+B_k'}{T_k} +
\sum_{i=1}^{k}\frac{C_i}{T_i} \leq 1,
\]
where
\begin{equation*}
  \label{eq:Bk}
  B_k = \sum_{i=1}^{k} \min\{S_i, C_i\}
\end{equation*}
\begin{equation*}
  \label{eq:Bkprime}
  B_k' = \max_{1 \leq i \leq k}\left(\max\{0, S_i - C_i\}\right).
\end{equation*}
\end{theorem}
Note that the notation follows the survey paper by
Chen~et~al.~\cite{Chen2018-suspension-review} instead of the original
paper by Devi~\cite{DBLP:conf/ecrts/Devi03}. Moreover, Devi considered
arbitrary-deadline task systems with asynchronous arrival times. Our
counterexample is valid by considering two implicit-deadline periodic
tasks released at the same time.

\section{Counterexample for Devi's Analysis}
\label{sec:counterexample}

The following task set $\textbf{T}$ with only two tasks provides a
counterexample for Devi's analysis:
\begin{itemize}
\item $\tau_1: (T_1=D_1=6, C_1=5, S_1=1)$ and
\item $\tau_2: (T_2=D_2=8, C_2=\epsilon, S_2=0)$, for any $0
  <\epsilon \leq 1/3$.
\end{itemize}

\begin{figure}[t]
\centering
\scalebox{0.8}{
	\begin{tikzpicture}[x=\ux,y=\uy,auto, thick]
    \draw[->] (0,0) -- coordinate (xaxis) (22,0) node[anchor=north]{$t$};
    \foreach \x in {0,1,...,20}{
		\draw[-,below](\x,0) -- (\x,-0.3) node[] {\pgfmathtruncatemacro\yi{\x} \yi};
	}
	\foreach \x in {0,1,...,20}{
         \draw[-,very thin,lightgray, dashed](\x,0.3) -- (\x,4);
	}	
	\foreach \y in {3.0}{
		\draw[] (0,\y) -- (22,\y);
	}

	\begin{scope}[shift={(0,3)}]
		\node[anchor=east] at (0, 0.5) {$\tau_1$};
                \draw[->, very thick] (0,0) -- (0,1.75);
		\foreach \x in {6,12,18}{ 
                  \draw[<->, very thick] (\x,0) -- (\x,1.75);
                }
                \node[task7, minimum width=\ux, anchor=south west] at (0, 0){};
                \node[task7, minimum width=4*\ux, anchor=south west] at (2, 0){};
                \node[task7, minimum width=\ux, anchor=south west] at (6, 0){};
                \node[task7, minimum width=4*\ux, anchor=south west] at (8, 0){};
                \node[task7, minimum width=\ux, anchor=south west] at (12.15, 0){};
                \node[task7, minimum width=4*\ux, anchor=south west] at (14.15, 0){};

        	\foreach \x in {1,7,13.15}{
        		\draw[] (\x,0) -- (\x,1);
        		\draw[] (\x+1,0) -- (\x+1,1);
                        \foreach \y in {0.3,0.5,0.7}{ 
        			\draw[] (\x,\y) -- (\x+1,\y);
                        }
        	}
                \draw[<-,red] (18,0) -- (18,1.49);
                \node[anchor=south,red] at (18, 1.6) {deadline miss};
				
	\end{scope}

	\begin{scope}[shift={(0,0)}]
		\node[anchor=east] at (0, 0.5) {$\tau_2$};
                \draw[<->, very thick] (0,0) -- (0,1.75);
		\foreach \x in {8,16}{ 
                  \draw[<->, very thick] (\x,0) -- (\x,1.75);
                }
		\foreach \x in {1, 12}{ 
    			\draw(\x, 0)  -- (\x, \uy) -- (\x+0.15, \uy) -- (\x+0.15, 0);
        	}        
	\end{scope}
      \end{tikzpicture}}
\caption{A concrete EDF schedule with a deadline miss.}
\label{fig:counter-example}
\end{figure}
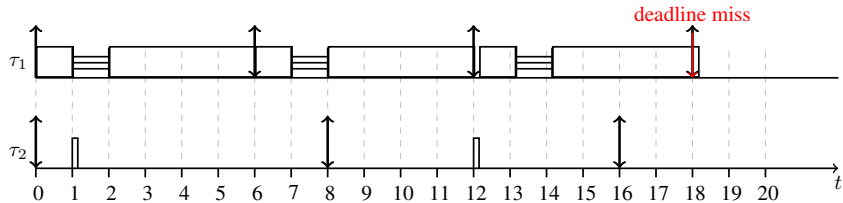

The test of Theorem~\ref{thm:devistheorem} is as follows:
\begin{itemize}
\item When $k=1$, we have $B_1 = 1$ and $B_1'=0$. Therefore, when
  $k=1$, $\frac{B_k+B_k'}{T_k} + \sum_{i=1}^{k}\frac{C_i}{T_i} = 1$.
\item When $k=2$, we have $B_2 = 1$ and $B_2'=0$. Therefore, when
  $k=2$, $\frac{B_k+B_k'}{T_k} + \sum_{i=1}^{k}\frac{C_i}{T_i} =
  \frac{1}{8} + \frac{\epsilon}{8} + \frac{5}{6} =
  \frac{23+3\epsilon}{24} \leq 1$, since $\epsilon \leq 1/3$.
\end{itemize}
Therefore, Devi's schedulability test concludes that the task set is
feasibly scheduled by preemptive EDF. But, a concrete schedule as
demonstrated in Figure~\ref{fig:counter-example} shows that one of the
jobs of task $\tau_1$ misses its deadline even when both tasks release
their first jobs at the same time.

The example in Figure~\ref{fig:counter-example} shows that a job of
task $\tau_1$ may be blocked by a job of task $\tau_2$, which results
in a deadline miss of the job of task $\tau_1$. However, in Devi's
schedulability analysis, such blocking is never considered since $B_1$
and $B_1'$ do not have any term related to $\tau_2$.

\section{Conclusion and Discussions}

The counterexample in Section~\ref{sec:counterexample} only requires
task $\tau_1$ to suspend once. It shows that applying
Devi's analysis in~\cite{DBLP:conf/ecrts/Devi03} is unsafe even for the
segmented self-suspension model under EDF scheduling.
We note that the above counterexample is only for
Theorem 8 in \cite{DBLP:conf/ecrts/Devi03}. We do not examine any
other schedulability tests in \cite{DBLP:conf/ecrts/Devi03}.

Although there have been many different analyses for preemptive fixed-priority scheduling, 
the only results for preemptive EDF are the analyses by
Liu and Anderson~\cite{DBLP:conf/ecrts/LiuA13}, and Dong and Liu~\cite{DBLP:conf/rtss/DongL16}, which are originally formulated for multiprocessor systems, the suspension-aware analysis by Devi, and the trivial suspension-oblivious analysis, which considers suspension time of the
self-suspending tasks as if they are usual execution time. (Detailed
discussions can be found in \cite[Section~4]{Chen2018-suspension-review}.)
The invalidation of Devi's analysis implies, that for preemptive EDF scheduling, there is no suspension-aware schedulability test specialized for uniprocessor systems.

\bibliography{real-time}{}

\end{document}